\documentclass[pra,twocolumn,amsmath,longbibliography]{revtex4-1}
\usepackage{graphicx}
\usepackage{amsmath}
\usepackage[dvipsnames]{xcolor}
\usepackage{soul}
\usepackage[normalem]{ulem}
\begin{document}

\title{Long-range states and Feshbach resonances in collisions between \\ ultracold alkali-metal diatomic molecules and atoms}

\author{Matthew D. Frye}

\affiliation{Joint Quantum Centre (JQC) Durham-Newcastle, Department of
Chemistry, Durham University, South Road, Durham DH1 3LE, United Kingdom}

\author{Jeremy M. Hutson}
\affiliation{Joint Quantum Centre (JQC) Durham-Newcastle, Department of
Chemistry, Durham University, South Road, Durham DH1 3LE, United Kingdom}

\date{\today}

\begin{abstract}
We consider the long-range states expected for complexes formed from an alkali-metal diatomic molecule in a singlet state and an alkali-metal atom. We explore the structure of the Hamiltonian for such systems, and the couplings between the six angular momenta that are present. We consider the patterns and densities of the long-range states, and the terms in the Hamiltonian that can cause Feshbach resonances when the states cross threshold as a function of magnetic field. We present a case study of $^{40}$K$^{87}$Rb+$^{87}$Rb. We show multiple types of resonance due to long-range states with rotational and/or hyperfine excitation, and consider the likelihood of them existing at low to moderate magnetic fields.
\end{abstract}

\maketitle

\section{Introduction}

Collisions are fundamentally important in ultracold physics. They not only dictate the lifetimes of ultracold samples and the efficiency of evaporative cooling, but also allow exquisite control of ultracold gases. For ultracold atoms, a detailed understanding of two-body collisions has evolved through close interplay between experiment and theory. In particular, a zero-energy Feshbach resonance occurs whenever a molecular bound state is close in energy to an atomic threshold, and is coupled to it by the interaction potential. A \emph{tunable} Feshbach resonance occurs when the state can be tuned across the threshold, most commonly with a magnetic field. At the lowest threshold, the s-wave scattering length passes through a pole as a function of applied field. This allows the effective interaction strength in a quantum gas to be tuned to any desired value in the vicinity of a resonance. As a result, magnetically tunable Feshbach resonances have become a mainstay of ultracold atomic physics \cite{Chin:RMP:2010}, with applications that range from the study of Efimov physics \cite{Naidon:2017} to investigations of the BCS-BEC crossover in degenerate Fermi gases \cite{Bloch:2012}. Feshbach resonances have also been used for magnetoassociation, in which pairs of ultracold atoms are converted to weakly bound diatomic molecules by sweeping a magnetic field across a resonance \cite{Regal:40K2:2003, Herbig:2003, Kohler:RMP:2006}. Several of these molecules have been transferred to the absolute ground state, usually by stimulated Raman adiabatic passage \cite{Ni:KRb:2008, Danzl:v73:2008, Lang:cruising:2008, Danzl:ground:2010,
Takekoshi:RbCs:2014, Molony:RbCs:2014, Park:NaK:2015, Guo:NaRb:2016, Rvachov:2017, Seesselberg:2018,
Yang:K_NaK:2019, Voges:NaK:2020, Cairncross:2021}.

Many new opportunities will open up if molecular collisions can be controlled in the same way as atomic collisions, through tunable Feshbach resonances. However, ultracold molecule-molecule collisions have turned out to be unexpectedly lossy \cite{Gregory:RbCs-collisions:2019, Bause:2021, Gersema:2021}, even in systems where there is no energetically allowed two-body reaction between the colliding species \cite{Zuchowski:trimers:2010}. Indeed, most experiments on diatomic molecules have observed loss rates close to the so-called universal rate, corresponding to unit probability of loss for colliding pairs that reach short range \cite{Idziaszek:PRL:2010}.
This has prompted work to engineer long-range potentials that prevent collisions reaching short range \cite{Wang:dipolar:2015, Karman:shielding:2018, Matsuda:2020, Anderegg:2021, Li:KRb-shield-3D:2021, Schindewolf:NaK-degen:2022}. The short-range loss is usually explained in terms of the formation of long-lived collision complexes \cite{Mayle:2012, Mayle:2013}, which cause a very dense (and so far unresolved) mass of Feshbach resonances. The complexes can be destroyed in a variety of ways, including by chemical reaction or via optical excitation by the trapping light \cite{Christianen:laser:2019}. However, there is conflicting evidence on the lifetimes of the complexes and the mechanisms of their loss \cite{Gregory:RbCs-complex-lifetime:2020, Liu:2020, Gregory:atom-molecule:2021, Bause:2021, Gersema:2021}.

Tunable Feshbach resonances have not yet been observed in molecule-molecule collisions. However,
ultracold atom-molecule collisions provide a middle ground between the simplicity of atom-atom collisions and the complexity of molecule-molecule collisions. Collisional losses are still fast in some cases, but much slower in others. The first observations of atom-molecule Feshbach resonances have recently been made. In particular, magnetic Feshbach resonances have been observed in collisions between ultracold $^{40}$K atoms and $^{23}$Na$^{40}$K molecules in singlet states \cite{Yang:K_NaK:2019, Wang:K_NaK:2021, Su:elastic:2022, Yang:rf-association:2022, Yang:magnetoassociation:2022} and between $^{23}$Na atoms and $^6$Li$^{23}$Na molecules in triplet states \cite{Son:Na_NaLi:2022}.

We have recently investigated the theory of the triatomic complexes that can be formed in collisions between alkali-metal diatomic molecules and atoms \cite{Frye:triatomic-complexes:2021}. We focussed on the doublet states formed in collisions of diatomic molecules in singlet states. We calculated densities of states for atom-molecule pairs at short range and showed that, for a single hyperfine manifold, the states are much more widely spaced than the range of collision energies. These states are expected to be chaotic in nature \cite{Mayle:2012, Frye:chaos:2016}. We also considered the role of hyperfine coupling due to the Fermi contact interaction, and showed that it can couple different hyperfine manifolds, producing broad resonances and background loss. However, the short-range states are unlikely to be responsible for the resonances observed in K+NaK; the patterns of the resonances observed suggest that they are due to long-range states \cite{Wang:K_NaK:2021}, which are more weakly coupled to the scattering channels and cause narrower resonances that are potentially controllable with magnetic fields.

\begin{figure}[tbp]
\includegraphics[width=0.99\linewidth]{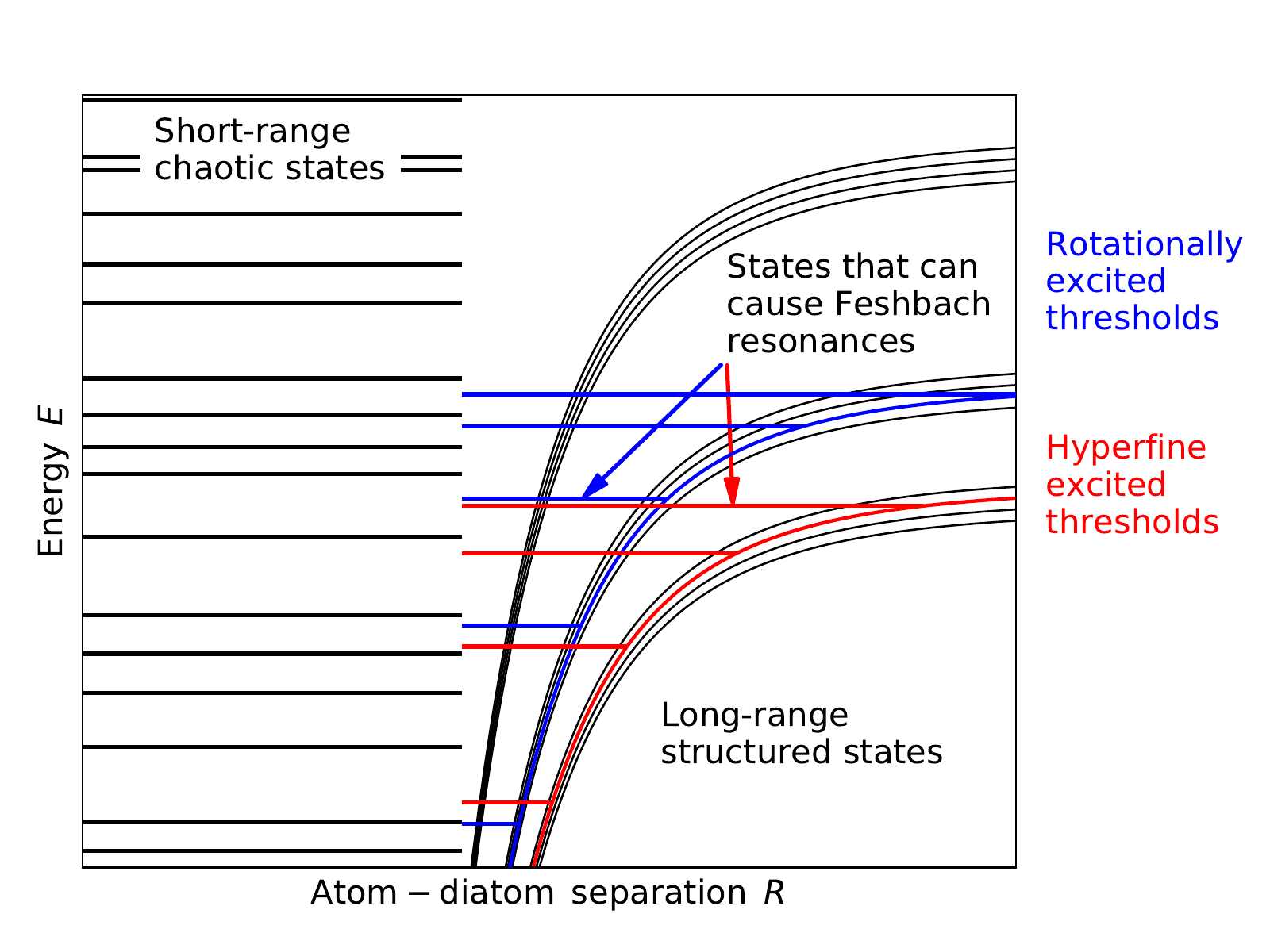}
\caption{\label{fig:schematic} Schematic picture of the channels and states in an alkali-metal atom+diatom collision. Short range states are expected to be chaotic, but the asymptotic channels have atom+diatom character, and so do any states supported by those channels in the long-range region.
Examples of long-range states are shown for one channel with rotational excitation (blue) and one channel with hyperfine excitation (red) but similar states will also exist for other channels.
States that lie close to the incoming channel threshold can cause magnetically tunable Feshbach resonances.}
\end{figure}

The purpose of the present paper is to investigate the long-range states of the complexes that can be formed in collisions between alkali-metal diatomic molecules and atoms, and the circumstances under which they can produce tunable Feshbach resonances. The long-range states are relatively weakly coupled to one another; they are likely to have more structured energy-level patterns than the short-range states, and not to be chaotic in nature. Nevertheless, they have complicated Zeeman and hyperfine Hamiltonians, with many terms that can couple them to the incoming and (in some cases) inelastic scattering channels. These states are shown schematically in Fig.~\ref{fig:schematic}.

The structure of the paper is as follows. Section \ref{sec:ham} explores the form of the Hamiltonian for atom-diatom complexes, the ways that different angular momenta are coupled together, and the selection rules rules for different coupling terms. Section \ref{sec:Feshbach} considers the patterns and densities of near-threshold states, and their consequences for Feshbach resonances in different alkali molecule-atom systems. Section \ref{sec:KRbRb} describes a case study for $^{40}$K$^{87}$Rb+$^{87}$Rb, with specific examples of level-crossing diagrams for resonances due to different types of long-range state. Finally, Section \ref{sec:conc} presents our conclusions.

\section{Hamiltonian and selection rules} \label{sec:ham}

There are 6 sources of angular momentum in a triatomic system AB+C formed from a singlet
molecule and an alkali-metal atom: the electron spin $S=1/2$, three nuclear spins $i_\textrm{A}$,
$i_\textrm{B}$, $i_\textrm{C}$, the rotation $n$ of the diatomic molecule AB and the rotation $L$ of AB and C about one another.

\subsection{Atomic and molecular states}

An alkali-metal atom C in a $^2$S state, with electron spin $S=1/2$ and nuclear spin $i_\textrm{C}$, is characterized in zero field by its total spin $f_\textrm{C}=i_\textrm{C}\pm\frac{1}{2}$. The two hyperfine states are separated by the hyperfine splitting $(i_\textrm{C}+\frac{1}{2}) \zeta_\textrm{C}$, where $\zeta_\textrm{C}$ is the scalar hyperfine coupling constant that arises from the Fermi contact interaction; the splitting varies from $228$ MHz for $^6$Li to 9.2 GHz for $^{133}$Cs. In a small magnetic field $B$, each hyperfine state $f_\textrm{C}$ is split into $2f_\textrm{C}+1$ Zeeman states labeled by the projection $m_{f,\textrm{C}}$ onto the axis of the field. At higher fields, the Zeeman effect mixes the two states of different $f_\textrm{C}$, and at very high fields the good quantum numbers are $M_S$ and $m_\textrm{C}$ rather than $f_\textrm{C}$, with $m_{f,\textrm{C}}=M_S+m_\textrm{C}$. The states at any field may nevertheless be labeled $(\tilde{f}_\textrm{C},m_{f,\textrm{C}})$, where $\tilde{f}_\textrm{C}$ is the value of $f_\textrm{C}$ that the state correlates with at zero field.

A diatomic molecule AB in a $^1\Sigma$ state, with vibrational and rotational quantum numbers $v$ and $n$, is characterized in zero field by its total angular momentum $f_\textrm{AB}$, which is the resultant of $n$, $i_\textrm{A}$ and $i_\textrm{B}$. For the ground rotational state, $n=0$, $i_\textrm{A}$ and $i_\textrm{B}$ are coupled only very weakly, with small splittings between the zero-field states with different values of $f_\textrm{AB}$. For the alkali-metal diatomic molecules, the spread is typically only tens of kHz and is due to the scalar nuclear-spin--nuclear-spin interaction \cite{Aldegunde:polar:2008, Aldegunde:singlet:2017}. Thus only a small magnetic field (tens of Gauss) is needed to decouple the nuclear spins \cite{Aldegunde:polar:2008, Aldegunde:singlet:2017}, such that the projections $m_\textrm{A}$ and $m_\textrm{B}$ of $i_\textrm{A}$ and $i_\textrm{B}$ become nearly good quantum numbers. However, for $n>0$ there are much larger zero-field splittings, usually dominated by nuclear quadrupole couplings, but including nuclear-spin--rotation and scalar and tensor nuclear-spin--nuclear-spin interactions; for these states a considerably larger magnetic field (hundreds of Gauss) is needed to decouple $i_\textrm{A}$ and $i_\textrm{B}$ from $n$ \cite{Ran:2010}, and $f_\textrm{AB}$ may remain a useful quantum number at considerably higher fields. In an optical trap, states with $n>0$ are further complicated by mixing due to the anisotropic ac Stark effect \cite{Gregory:RbCs-AC-Stark:2017}.

\subsection{The interaction operator}

Colliding atoms and molecules are initially in eigenstates of their individual Hamiltonians. Terms that are present in the Hamiltonians of the separated species, and are independent of their separation $R$, do not provide coupling between the colliding pair and the bound states of the triatomic complex. They therefore cannot by themselves produce Feshbach resonances. It is convenient to define an \emph{interaction operator} $\hat{V}(R,\xi)$ that includes all $R$-dependent terms in the Hamiltonian of the colliding pair, excluding the relative kinetic energy. Here $\xi$ represents all coordinates except $R$, including spin coordinates. The interaction operator includes the intermolecular potential, but also contains spin-dependent and field-dependent terms as described below.

The complete wavefunction of the collision system or the triatomic complex may be expanded
\begin{equation} \Psi(R,\xi)
=R^{-1} \sum_j \Phi_j(\xi) \psi_{j}(R). \label{eq:exp}
\end{equation}
The \emph{channel functions} $\Phi_j(\xi)$ form a complete orthonormal basis set for motion in the coordinates $\xi$.
The component of the wavefunction in each channel $j$ is described by a \emph{radial channel function} $\psi_{j}(R)$. A full numerical solution of the resulting coupled-channel problem is beyond the scope of this paper, but the formalism is nonetheless useful for considering the various couplings and their effects.

When considering collisions and long-range states, it is convenient to choose channel functions with quantum numbers $\{v,n,L,N,M_N,m_\textrm{A},m_\textrm{B},\tilde{f}_\textrm{C},m_{f,\textrm{C}}\}$, collectively represented by $j$ in Eq.\ (\ref{eq:exp}). Here $N$ is the total spin-free angular momentum, which is the resultant of $n$ and $L$, and $M_N$ is its projection onto the axis of the magnetic field. The channel functions are defined more formally in Appendix A. It is useful to distinguish a quantum number for the incoming channel, $X^\textrm{in}$, from the corresponding one for a resonant bound state, $X^\textrm{res}$.

For each term $\hat{\Omega}$ in the interaction operator, we are interested in the matrix elements
$\langle j^\textrm{res} | \hat{\Omega} | j^\textrm{in} \rangle$
and particularly in the associated selection rules $\Delta X=X^\textrm{res}-X^\textrm{in}$ involving the different quantum numbers described above.
We now consider these selection rules for each term in the interaction operator.

\subsubsection{Interaction terms diagonal in $N$}

The strongest term in the interaction operator is the electrostatic interaction potential $V(R,r,\theta)$. This is written here in Jacobi coordinates: $\vec{r}$ is the interatomic vector of AB, of length $r$, $\vec{R}$ is the vector from the center of mass of AB to atom C, of length $R$, and $\theta$ is the angle between $\vec{r}$ and $\vec{R}$.
For the alkali-metal systems of interest here, $V(R,r,\theta)$ is deep (of order 50~THz) and highly anisotropic at short range. Such potentials have been considered in depth elsewhere
\cite{Martin:1978, Varandas:1982, Martins:1983, Varandas:1986, vonBusch:1998, Hauser:2008, Zuchowski:trimers:2010, Hauser:2010, Hauser:2015, Croft:K+KRb:2017, Kendrick:Li+LiNa:2021}, but their details are not crucial for the present purposes. The important feature of the short-range potential is that it
provides strong coupling between channels with different values of $v$, $n$ and $L$.
However, it is diagonal in $N$ and its projection $M_N$. It is also diagonal in all spin quantum numbers. Ultracold collisions are usually dominated by the incoming s wave, with $L^\textrm{in}=0$, which implies $N^\textrm{in}=n^\textrm{in}$. If the molecule AB is initially in $n^\textrm{in}=0$, the electrostatic potential couples the incoming wave only to other channels with $N^\textrm{res}=0$, requiring $L^\textrm{res}=n^\textrm{res}$, and the same spin quantum numbers as the incoming channel.

There are also terms in the interaction operator that depend on spins and/or applied fields. All the hyperfine and Zeeman terms that exist in the separated atom and molecule become functions of $R$ and $\theta$ when the two species interact, and the differences from the monomer terms contribute to the interaction operator. In addition, there are scalar and tensor interactions that develop between the angular momenta on the atom and those on the molecule. The magnetic moments are also modified, and contribute Zeeman terms to the interaction operator.

The strongest spin-dependent term is the Fermi contact interaction. At long range this couples $S$ to $i_\textrm{C}$ for the free atom to form $f_\textrm{C}$. We showed in Ref.\ \cite{Frye:triatomic-complexes:2021} that the Fermi contact interaction depends strongly on the geometry of the complex, particularly at short range, where the electron spin that is originally on atom C is distributed among all three atoms. The Fermi contact operator is of the form \cite{Frye:triatomic-complexes:2021}
\begin{equation}
\hat{H}_\textrm{Fc}=\sum_{X=\textrm{A,B,C}} \zeta_X(R,r,\theta) \hat{\boldsymbol{i}}_X\cdot\hat{\boldsymbol{S}}, \label{eq:fermi}
\end{equation}
where $\hat{\boldsymbol{S}}$ and $\hat{\boldsymbol{i}}_X$ are the vector operators for the electron and
nuclear spin angular momenta.
Each term is the product of a scalar spatial operator $\zeta_X(R,r,\theta)$, which can couple states of different $v$, $n$ and $L$ while conserving $N$ and $M_N$, and a spin operator $\hat{\boldsymbol{i}}_X\cdot\hat{\boldsymbol{S}}$. The latter provides a coupling that can change $\tilde{f}_\textrm{C}$ and/or $m_{f,\textrm{C}}$ by 0 or 1, while conserving the total spin projection $m_{f,\textrm{tot}}=m_\textrm{A}+m_\textrm{B}+m_{f,\textrm{C}}$.

If all the external fields present share a common axis, the total projection quantum number $M_F$ is conserved,
\begin{equation}
M_F = M_N + m_{f,\textrm{tot}} = M_N + M_S + m_\textrm{A} + m_\textrm{B} + m_\textrm{C}.
\end{equation}
Conservation of $m_{f,\textrm{tot}}$ thus implies conservation of $M_N$, and \emph{vice versa}.

\subsubsection{Interaction terms off-diagonal in $N$} \label{sec:off_diag}

An important question is the strength of the coupling between channels with different values of $N$ and $M_N$. If such couplings are significant, bound states with one value of $(N,M_N)$ may cause Feshbach resonances in an incoming channel with different values. In experiments that involve diatomic molecules in their rotationless ground state, $n^\textrm{in}=0$, and are dominated by s-wave scattering, $L^\textrm{in}=0$, the incoming channel has $N^\textrm{in}=0$. Couplings off-diagonal in $N$ are needed for collision complexes in any state $(N^\textrm{res},M^\textrm{res}_N)$ other than (0,0) to contribute in such cases. 

There are tensor hyperfine terms in the Hamiltonian, arising from dipolar interactions between electron and nuclear spins. These terms are off-diagonal in $m_X$ for the nucleus concerned by up to 1, in $M_S$ by up to 1, and in $n$, $L$, $N$ and $M_N$ by up to 2, while conserving $M_F$. They depend on the asymmetry of the spin density around the nucleus concerned. The tensor coupling constants for Na$_3$ at its equilibrium geometry have been obtained from electronic structure calculations \cite{Hauser:2015} and found to be up to 142~MHz. They are zero for free atoms and singlet molecules, so form part of the interaction operator. The calculated value for Na$_3$ is about 16\% of the
atomic hyperfine coupling of Na; it seems reasonable to expect similar percentages for the triatomic complexes of interest here.

There are additional hyperfine terms arising from the interaction of nuclear electric quadrupole moments with the electric field gradient at the nucleus concerned. These too are off-diagonal in $n$, $L$, $N$ and $M_N$ by up to 2. However, the coupling constants are generally less than 10 MHz for alkali-metal diatomic molecules \cite{Aldegunde:singlet:2017}, and are probably comparable in the 3-atom complexes. They are again zero for free atoms, so there is some dependence on $R$, but their absolute magnitude is small.

A magnetic field $B$ interacts mostly with the magnetic moment due to the electron spin $S$ and to a much lesser extent with those due to the nuclear spins $i_X$ and the rotational angular momenta $n$, $L$ and $N$. It is likely that only the electron-spin contribution is significant in the interaction operator. For a free atom, the matrix elements are of magnitude $g_S \mu_\textrm{B} M_S B$, where $g_S$ is the g-factor for the electron spin and $\mu_\textrm{B}$ is the Bohr magneton. The electron g-tensor for a molecule is anisotropic and geometry-dependent, with components $g_{ij}$ in a frame fixed in the molecule.
The term that enters the interaction operator depends on the deviation of $g_{ij}(R,r,\theta)$ from $g_S$; such deviations are small (usually less than 1\%) for molecules with well-separated electronic states that contain only light atoms, but they may be enhanced if there is a nearby excited state with a spin density spatially close to that of the state of interest \cite{Lushington:1997}. For example, the shifts can be a substantial fraction of $g_S$ for transition-metal species \cite{Abragam:1970}. Moreover, the spin-orbit matrix elements responsible for them scale roughly as $Z^4$ with nuclear charge $Z$. They might therefore be substantial for the alkali-metal triatomic molecules. They are diagonal in $M_S$ and therefore off-diagonal in $\tilde{f}_\textrm{C}$ and $m_{f,\textrm{C}}$. Since the g-tensor is defined in a frame fixed in the triatomic complex, it must be rotated to evaluate its matrix elements between space-fixed functions; as a result its anisotropic part can mix states with different $N$ and $M_N$.

There are also spin-rotation terms arising from the interaction of the electron spin with molecular
rotation and internal rotation. These arise mainly from the combined effects of spin-orbit and
Coriolis mixing between electronic states \cite{VanVleck:RMP:1951}. The spin-rotation coupling
tensor is approximately related to the g-tensor through Curl's approximation \cite{Curl:1965},
\begin{equation}
g_{ij} = g_S \delta_{ij} - \hbar^{-2} \sum_k \epsilon_{ik} I_{kj}.
\end{equation}
Here $I_{kj}$ is an element of the molecular inertial tensor, so that the spin-rotation constants
scale roughly with rotational constants as well as g-tensor shifts. The isotropic part has matrix
elements off-diagonal in $M_N$, while the anisotropic part is off-diagonal in $N$ and $M_N$; both
parts conserve $M_J=M_N+M_S$. Such terms are known to be up to several GHz for the ground states of
molecules such as NF$_2$, NO$_2$ and ClO$_2$ \cite{Curl:1965, Tarczay:2010}, but values below 1~MHz
have been obtained from electronic structure calculations on Na$_3$ at near-equilibrium geometries
\cite{Hauser:2015}.

An electric field interacts strongly with an electric dipole moment to produce matrix elements
diagonal and off-diagonal in $N$. However, there are no purposely applied electric fields in the
experiments considered here.

In an optical trap, the light from the trapping laser can interact with the molecular polarizability through the ac Stark effect \cite{Gregory:RbCs-AC-Stark:2017}. The polarizability is a second-rank tensor that depends on the geometry of the triatomic complex, so the AC Stark effect contributes to the interaction operator. It has matrix elements that can change $N$ by 0 or 2 and can also be off-diagonal in $v$, $n$ and $L$. It is diagonal in the spin quantum numbers.

\subsubsection{Calculation of magnetic properties}

The Fermi contact interactions, tensor hyperfine couplings, spin-rotation couplings and g-tensor components can in principle be obtained from electronic structure calculations. For simple species, considerable success has been achieved both with wavefunction-based methods \cite{Gauss:2009, Datta:2019} and with density-functional theory \cite{vanLenthe:1998, Neese:2001, Neese:2005, Autschbach:2011}. However, for the alkali-metal 3-atom complexes, the task is complicated by the existence of two low-lying electronic states, with seams of conical intersections between them as described in Ref.\ \cite{Frye:triatomic-complexes:2021}. Hauser \emph{et al.}\ \cite{Hauser:2015} have used a variety of electronic structure methods to calculate some of these properties in Na$_3$. However, they focused on high-symmetry geometries around the path sampled by pseudorotation in the lowest vibronic state, and obtained coupling constants involving spin densities by a fairly coarse integration over electronic coordinates. Extending their calculations to heteronuclear systems at a wide range of geometries is beyond the scope of the present paper.

\section{Long-range states and Feshbach resonances
\label{sec:Feshbach}}

\subsection{Structure of long-range states}

The states of complexes formed in alkali-metal atom-diatom collisions may be loosely separated into short-range and long-range states. The short-range states are supported by high-lying closed channels, with large values of $n$ and/or $v$. They spend most of their time in regions where the interaction potential $V(R,r,\theta)$ is deep and strongly anisotropic. Such states are strongly coupled to one another and are likely to be chaotic \cite{Mayle:2012, Frye:chaos:2016}. The resulting resonances were investigated in Ref.\ \cite{Frye:triatomic-complexes:2021}: the states may have magnetic moments significantly different from the colliding atom and molecule, so the resonances are magnetically tunable, but the coupling between the bound states and the continuum is very strong; for atom-diatom systems in which the levels are chaotic the resulting resonances are generally very broad (typically hundreds of Gauss).

Interaction potentials with asymptotic form $-C_j/R^j$ support states with wavefunctions concentrated at long range \cite{LeRoy:1970}. The spatial overlap between the long-range and short-range states decreases as threshold is approached. The long-range states are thus relatively weakly coupled to the short-range states by the potential anisotropy and other short-range interactions; the longest-range states are likely to be separate from the chaotic manifold, though they will exhibit avoided crossings with short-range states as a function of magnetic field. States of this general character have been observed experimentally in the chaotic spectrum of dysprosium dimers \cite{Maier:universal:2015} and also appear close to threshold in calculations on the chaotic system Li + CaH \cite{Frye:chaos:2016}. This picture contrasts with that of Mayle \emph{et al.}\ \cite{Mayle:2012}, who assumed complete statistical mixing for all states without distinction.

The long-range states are approximately described by the quantum numbers of the free atom C and molecule AB. To a first approximation, the long-range states of the complex are characterized by $\tilde{f}_\textrm{C}$ and $m_{f,\textrm{C}}$, together with the rotational quantum number $n$ of the diatomic molecule AB and a vibrational quantum number $\eta$ for motion in the atom-molecule separation $R$. For present purposes $\eta$ is most conveniently counted downwards from threshold, such that the least-bound state in each channel with $L=0$ is $\eta=-1$. The hyperfine and Zeeman structure of the alkali-metal diatomic molecule is complicated \cite{Aldegunde:polar:2008}, particularly in the presence of strong laser fields \cite{Gregory:RbCs-AC-Stark:2017}, but in general $n$ couples relatively weakly to the nuclear spins $i_\textrm{A}$ and $i_\textrm{B}$ with hyperfine and Zeeman splittings of a few MHz or less.

For long-range complexes at zero field, $n$ couples to $L$ to form resultant $N$, which couples to $f_\textrm{C}$ to form resultant $F_\textrm{C}$; this couples to $i_\textrm{A}$ and $i_\textrm{B}$ to form the total angular momentum $F$. However, long-range states of the same $(n,L,N)$ and $f_\textrm{C}$ but different $F$ are unlikely to be split by more than a few MHz, except in the vicinity of avoided crossings with short-range states. It therefore takes a magnetic field of only a few G to decouple $M_N$ and the nuclear spins of the diatomic molecule, so that the quantum numbers are $(n,L,N,M_N,\tilde{f}_\textrm{C},m_{f,\textrm{C}},m_\textrm{A},m_\textrm{B})$. The Zeeman effects for these states are dominated by that of atom C, and at fields more than a few G they form groups, spaced by no more than a few MHz, with the same values of $(n,L,N,\tilde{f}_\textrm{C},m_{f,\textrm{C}})$ but differing $(M_N,m_\textrm{A},m_\textrm{B})$. The total projection $M_F=M_N+m_{f,\textrm{C}}+m_\textrm{A}+m_\textrm{B}$ is conserved if the trapping laser is polarized parallel to the magnetic field.

For slightly deeper states with $N>0$, the anisotropy $V_\textrm{aniso}$ of the interaction potential $V(R,r,\theta)$ may be sufficient to quantize $n$ along the intermolecular vector with projection $K$. In this case $(n,L,N)$ is replaced by $(n,K,N)$. This recoupling occurs when $V_\textrm{aniso}$ is large compared to the rotational constant of the complex, $\hbar^2/(2\mu R^2)$, in the same way as for Case 2 coupling in spin-free Van der Waals complexes \cite{Hutson:AMVCD:1991}. For K+NaK, for example, it takes place near $R \sim 100\,a_0$.
As the anisotropy increases further at shorter range, $V_\textrm{aniso}(R)$ becomes greater than the rotational constant of the diatomic molecule, $\hbar^2/(2\mu_\textrm{AB}r^2)$. The diatom rotational quantum number $n$ is then replaced by a quantum number for a bending vibration, as for Case 3 coupling in Van der Waals complexes with strong anisotropy \cite{Hutson:AMVCD:1991}. Eventually, as the binding energy increases, the couplings to the short-range states will become strong enough that the long-range states will merge into the chaotic manifold, with level spacings described by random-matrix theory as in Ref.\ \cite{Frye:triatomic-complexes:2021}.

Feshbach resonances require coupling between the incoming scattering state and a quasibound state. Although the quantum numbers $(n,L,N,M_N, \tilde{f}_\textrm{C},m_{f,\textrm{C}}, m_\textrm{A},m_\textrm{B})$ described above are \emph{approximately} conserved for the long-range states, they are not \emph{fully} conserved. The long-range states do mix with the short-range states, and the mixed states have a component at short range; the couplings described in Section \ref{sec:ham} then provide the necessary coupling to the incoming state. In particular, magnetically tunable Feshbach resonances occur when a bound or quasibound state crosses the energy of the incoming state; they can be observed only if the incoming state and the quasibound state have different magnetic moments, and this will usually require differing values of $\tilde{f}_\textrm{C}$ or $m_{f,\textrm{C}}$; since $M_F$ is conserved, a change in $m_{f,\textrm{C}}$ requires a compensating change in $M_N$, $m_\textrm{A}$ or $m_\textrm{B}$. As described in Section \ref{sec:ham}, the Fermi contact interaction can exchange angular momentum between $m_{f,\textrm{C}}$ and $m_\textrm{A}$ or $m_\textrm{B}$ while conserving $m_{f,\textrm{tot}}$, with selection rule $\Delta m_{f,\textrm{C}}=\pm1$. It can also change $\tilde{f}_\textrm{C}$. Couplings off-diagonal in $N$ and/or $M_N$ are generally weaker but can arise from tensor hyperfine coupling, g-tensor anisotropy and spin-rotation coupling as described above.

The alkali molecule + atom systems have some similarities with systems in which an alkali-metal atom interacts with a closed-shell atom, such as Rb+Sr \cite{Zuchowski:RbSr:2010, Barbe:RbSr:2018} and alkali+Yb \cite{Brue:AlkYb:2013, Yang:CsYb:2019}. In both cases the resonances are principally due to $R$-dependent hyperfine coupling, and for Sr and Yb they are very narrow, with widths in the mG range. However, the alkali triatomic systems differ because the near-threshold states typically have much larger components at short range, due to mixing with the chaotic manifold of short-range states. In consequence, the resulting resonances will be substantially wider.

The long-range states will be much less susceptible to laser-induced loss than short-range states. For large $R$, the electronically excited triatomic states responsible for loss are much less strongly bound, so are likely to be inaccessible with the laser frequencies used for trapping. In an alternative viewpoint, the vibronically excited states responsible for loss are short-range states, which have poor Franck-Condon overlap with the long-range states. Because of this, the long-range states can be sharp enough to cause well-defined Feshbach resonances \cite{Yang:K_NaK:2019, Wang:K_NaK:2021}.

\begin{table*}[tbp]

\caption{Energy and length scales and bin depths for of $AB+A$ alkali-metal systems.
\label{Table:bins}} \centering
\begin{ruledtabular}
\begin{tabular}{c|cccccc}
System	& $\bar{a}$ (\AA)	& $\bar{E} / h$ (MHz)	& $E^{L=0}_\textrm{bin} / h$ (MHz)	& $E^{L=0}_\textrm{50\%} / h$ (MHz)	& $E^{L=1}_\textrm{bin} / h$ (MHz)	& $E^{L=2}_\textrm{bin} / h$ (MHz)	 \\ \hline
 $^{7}$Li$^{7}$Li+$^{7}$Li &   21 &  245 & 8861 & 1505 & 15150 & 22909 \\
 $^{7}$Li$^{23}$Na+$^{7}$Li & 20.8 &  205 & 7390 & 1255 & 12634 & 19104 \\
 $^{7}$Li$^{39}$K+$^{7}$Li & 23.2 &  154 & 5579 &  947 & 9539 & 14424 \\
 $^{7}$Li$^{87}$Rb+$^{7}$Li & 24.2 &  132 & 4784 &  812 & 8179 & 12368 \\
 $^{7}$Li$^{133}$Cs+$^{7}$Li & 25.4 &  117 & 4226 &  718 & 7224 & 10924 \\
 $^{7}$Li$^{23}$Na+$^{23}$Na &   26 & 57.3 & 2069 &  351 & 3537 & 5349 \\
 $^{23}$Na$^{23}$Na+$^{23}$Na &   29 & 39.1 & 1411 &  240 & 2413 & 3649 \\
 $^{23}$Na$^{39}$K+$^{23}$Na & 30.8 & 31.8 & 1148 &  195 & 1963 & 2968 \\
 $^{23}$Na$^{87}$Rb+$^{23}$Na & 32.5 & 25.1 &  907 &  154 & 1551 & 2345 \\
 $^{23}$Na$^{133}$Cs+$^{23}$Na & 34.3 & 21.4 &  774 &  131 & 1324 & 2001 \\
 $^{7}$Li$^{39}$K+$^{39}$K &   36 & 18.5 &  667 &  113 & 1141 & 1725 \\
 $^{23}$Na$^{39}$K+$^{39}$K & 37.8 & 14.8 &  535 & 90.8 &  915 & 1383 \\
 $^{39}$K$^{39}$K+$^{39}$K & 41.7 & 11.2 &  404 & 68.6 &  691 & 1045 \\
 $^{39}$K$^{87}$Rb+$^{39}$K & 42.6 & 9.34 &  337 & 57.3 &  577 &  872 \\
 $^{39}$K$^{133}$Cs+$^{39}$K & 44.8 & 7.94 &  287 & 48.7 &  490 &  741 \\
 $^{7}$Li$^{87}$Rb+$^{87}$Rb & 45.7 & 5.36 &  194 & 32.9 &  331 &  500 \\
 $^{23}$Na$^{87}$Rb+$^{87}$Rb & 47.3 & 4.66 &  168 & 28.6 &  288 &  435 \\
 $^{39}$K$^{87}$Rb+$^{87}$Rb &   50 & 3.93 &  142 & 24.1 &  242 &  367 \\
 $^{87}$Rb$^{87}$Rb+$^{87}$Rb & 53.4 & 3.06 &  111 & 18.8 &  189 &  286 \\
 $^{87}$Rb$^{133}$Cs+$^{87}$Rb &   55 & 2.68 & 96.9 & 16.5 &  166 &  251 \\
 $^{7}$Li$^{133}$Cs+$^{133}$Cs & 55.7 & 2.39 & 86.3 & 14.7 &  148 &  223 \\
 $^{23}$Na$^{133}$Cs+$^{133}$Cs &   57 & 2.17 & 78.2 & 13.3 &  134 &  202 \\
 $^{39}$K$^{133}$Cs+$^{133}$Cs & 59.5 &  1.9 & 68.7 & 11.7 &  117 &  178 \\
 $^{87}$Rb$^{133}$Cs+$^{133}$Cs & 61.9 & 1.59 & 57.5 & 9.76 & 98.3 &  149 \\
 $^{133}$Cs$^{133}$Cs+$^{133}$Cs & 65.2 & 1.34 & 48.4 & 8.22 & 82.8 &  125 \\
 \end{tabular}
\end{ruledtabular}
\end{table*}

\subsection{Expected positions of Feshbach resonances}

\label{sec:bins}

The alkali-metal atom-molecule systems have interaction potentials of the form $V(R)=-C_6/R^6$ at long range. If channel coupling is neglected, each asymptotic channel labeled by quantum numbers $(n,L,N,M_N,\tilde{f}_\textrm{C},m_{f,\textrm{C}},m_\textrm{A},m_\textrm{B})$ supports a set of near-threshold bound states with a simple pattern of binding energies given by quantum defect theory \cite{Gao:2000}. There is always exactly one s-wave bound state ($L=0$) in a certain energy window immediately below threshold. This window is known as the top bin and its width is determined by only the asymptotic form of the interaction potential and the reduced mass $\mu$. For an interaction potential $-C_6R^{-6}$, the width of the top bin is approximately $36\bar{E}$, where $\bar{E}=\hbar^2/(2\mu\bar{a}^2)$ is the energy scale associated with the mean scattering length of Gribakin and Flambaum \cite{Gribakin:1993}, $\bar{a}=(2\mu C_6/\hbar^2)^{1/4}\times 0.4779888\dots$. Values of $\bar{a}$ and $\bar{E}$ are given in Table \ref{Table:bins} for all possible systems $AB+A$, where $A$ and $B$ are different alkali-metal atoms, using the $C_6$ coefficients of \cite{Zuchowski:vdW:2013}. The concept of bins can be extended to higher partial waves and deeper bound states, as described by Gao \cite{Gao:2000}. Table \ref{Table:bins} includes the depth of the top bin, $E^L_\textrm{bin}$, for $L=0$, 1 and 2.

To a first approximation, a near-threshold bound state retains the character of the threshold that supports it. If channel mixing is neglected, it shares the magnetic moment of the threshold and runs below it and parallel to it as a function of magnetic field.
Consider incoming and resonant channels with threshold energies $E^\textrm{in}(B)$ and $E^\textrm{res}(B)$ as a function of magnetic field. A bound state with binding energy $E^\textrm{res}_\textrm{b}$ relative to $E^\textrm{res}(B)$ crosses the incoming channel and may cause a Feshbach resonance near the field where $E^\textrm{in}(B)=E^\textrm{res}(B)-E^\textrm{res}_\textrm{b}$.
In the simple case where the Zeeman effect is linear, as for heavier alkali-metal atoms at low fields, and the channels correlate with the same zero-field level,
$E^\textrm{res}(B)-E^\textrm{in}(B)=-g_S \mu_\textrm{B} B \Delta m_{f,\textrm{C}}/(i_\textrm{C}+1/2)$.

The positions of states within their bins are governed by the scattering length $a$ and may be obtained from quantum defect theory \cite{Gao:QDT:1998, Gao:C6:1998, Gao:2001, Gao:AQDTroutines}.
An s-wave bound state exists close to the top of the bin if the scattering length is large and positive, $a\gg\bar{a}$. Similarly, a p-wave state ($L=1$) exists just below threshold if $a$ is slightly less than $2\bar{a}$ and a d-wave state ($L=2$) exists just below threshold if $a$ is slightly less than $\bar{a}$. For a potential with unknown scattering length, it is possible to calculate the \emph{probability} that a state with $L=0$, 1 or 2 exists in a particular range of binding energies. For example, there is a 50\% probability of an s-wave state bound by less than $E^{L=0}_\textrm{50\%}=6.1\bar{E}$; these values are included in Table~\ref{Table:bins}.

\begin{figure}[tbp]
\includegraphics[width=0.99\linewidth]{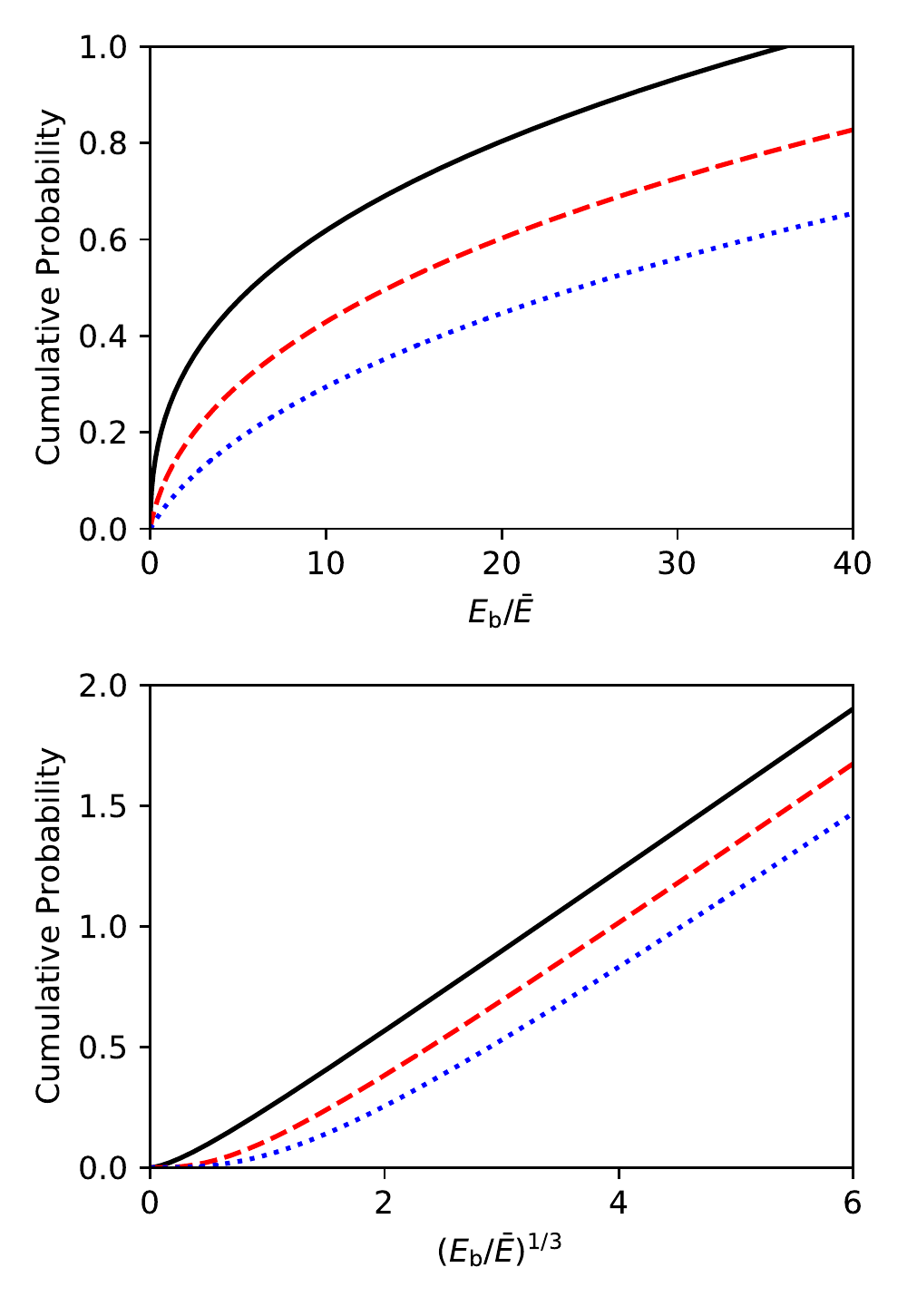}
\caption{\label{fig:bins} Cumulative probability of a bound state within energy $E$ of threshold for $L=0$ (black solid line), 1 (red dashed line), and 2 (blue dotted line). Panel (a) shows this on a linear energy scale to the depth of the first bin, and panel (b) shows it as a function of $(E/\bar{E})^{1/3}$ to larger binding energies. Probabilities above 1 indicate that there is at least one state, in the top bin, and a probability of a second state.}
\end{figure}

Fig.\ \ref{fig:bins}(a) shows the cumulative probability of a state with $L=0$, 1 or 2 existing within
an energy range $E_\textrm{b}/\bar{E}$ below threshold. Fig.\ \ref{fig:bins}(b) show the same quantity as a
function of $(E_\textrm{b}/\bar{E})^{1/3}$, which approaches linearity away from threshold. This may be
interpreted as the probability of a bound state of this $L$ causing a
resonance in an incoming channel at a field \emph{below} that where the differential Zeeman effect,
$E^\textrm{res}(B)-E^\textrm{in}(B)$, is equal to $E^\textrm{res}_\textrm{b}$. To estimate the probability of a resonance occurring below a
field $B$, calculate the differential Zeeman shift between the two atomic states, divide by
$\bar{E}$ from Table \ref{Table:bins}, and read off the probabilities.

When the couplings that produce resonances are short-range in character, the widths of
resonances are proportional to $E_\textrm{b}^{2/3}$ \cite{Brue:AlkYb:2013}. States very close to
threshold may therefore produce resonances that are too narrow to observe. Under these
circumstances the lowest-field resonances that are actually observable may be those at the top of
the second bin, at slightly higher fields than implied by Table \ref{Table:bins}. This was the
case, for example, for analogous resonances for $\tilde{f}_\textrm{Rb}=1$ in $^{87}$Rb+$^{87}$Sr
\cite{Barbe:RbSr:2018}.

\section{Case study: Long-range states of KR\lowercase{b}+R\lowercase{b}}
\label{sec:KRbRb}

\begin{figure*}[htb]
\includegraphics[width=0.98\linewidth]{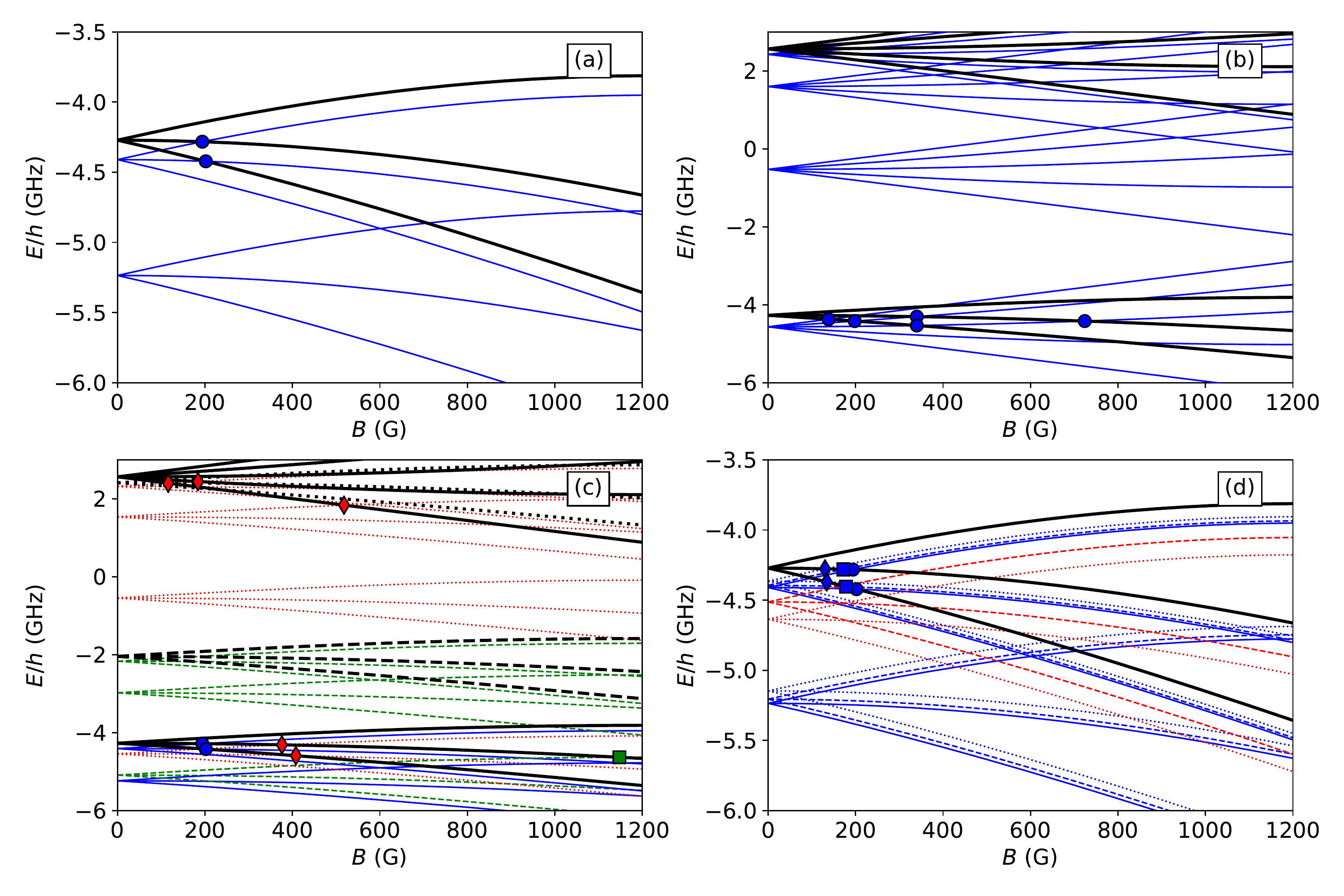}
\caption{\label{fig:crossing} Example level crossing diagrams for different types of resonances due to long-range states for $^{40}$K$^{87}$Rb+$^{87}$Rb. Heavy black lines show thresholds, and thin colored lines show bound states. Crossings that can cause Feshbach resonances are marked by symbols. All examples show single-channel bound states for $a=\infty$, for which there is a state at the bottom of each s-wave bin.
(a) States $(n^\textrm{res},\tilde{f}^\textrm{res}_\textrm{C},L^\textrm{res})=(0,1,0)$.
(b) States $(0,2,0)$.
(c) States $(0,1,0)$ are shown with blue solid lines and circles, (1,1,1) with green dashed lines and squares, and (2,1,2) with red dotted lines and diamonds.
(d) States $(0,1,0)$ are shown with solid lines and circles, (0,1,1) with dashed lines and squares, and (0,1,2) with dotted lines and diamonds. Also shown in red are the bottom of the bins for $L^\textrm{res}=1$ and 2.}
\end{figure*}

The bound states that cause Feshbach resonances in an incoming channel labeled by $(n^\textrm{in}=0,L^\textrm{in}=0,\tilde{f}^\textrm{in}_\textrm{C},m^\textrm{in}_{f,\textrm{C}})$ may belong to parent closed channels with differing values of $n^\textrm{res}$, $\tilde{f}^\textrm{res}_\textrm{C}$ or $m^\textrm{res}_{f,\textrm{C}}$ (or any combination of them). It is useful to consider a specific example here, and we choose $^{40}$K$^{87}$Rb+$^{87}$Rb with the Rb atom in its lowest state $(\tilde{f}^\textrm{in}_\textrm{C},m^\textrm{in}_{f,\textrm{C}})=(1,1)$ and the KRb molecule in the excited Zeeman state $(n^\textrm{in},m^\textrm{in}_\textrm{A},m^\textrm{in}_\textrm{B})=(0,-4,1/2)$.

Closed channels with $(n^\textrm{res},\tilde{f}^\textrm{res}_\textrm{C})=(0,1)$ are almost degenerate with the incoming channel at zero field; each such channel must have a least-bound state with $\eta=-1$ somewhere in the top 142 MHz. The Fermi contact term $\hat{H}_\textrm{Fc}$ has selection rule $\Delta m_{f,\textrm{C}}=0,\pm1$, but only $-1$ is relevant for this set of resonances. Since the differential magnetic moment between adjacent Zeeman states for Rb$(f_\textrm{C}=1)$ is 0.7~MHz/G at low field, the corresponding resonances are expected at fields below 200 G, depending on where the least-bound state lies within its bin. We show an illustrative example of a level-crossing diagram for these resonances in Fig.\ \ref{fig:crossing}(a). This is for the case $a=\infty$, for which there is a state in each channel at the bottom of each bin; the binding energies are again calculated using quantum defect theory. Corresponding examples for other values of $a$ are given in Supplemental Material \cite{supp:long-range}. The crossings for $a=\infty$ are not predictions of specific resonance positions, but instead predict a resonance due to a state with $\eta=-1$ somewhere between $B=0$ and the crossing marked at 200 G. There will also be a bound state in the second bin, with $\eta=-2$. This will lie between the two zero-field levels shown in Fig.\ \ref{fig:crossing}(a). It will cause a resonance above 200 G; the upper limit is more complicated due to non-linear Zeeman effects, and depends on $m_{f,\textrm{C}}$, but is readily calculated from the depth of bottom of the second bin, which is 980 MHz for this system. These resonances are analogous to Type II resonances that can occur in systems such as LiYb, RbSr and CsYb \cite{Brue:LiYb:2012, Barbe:RbSr:2018, Yang:CsYb:2019}.

The change $\Delta m_{f,\textrm{C}}=-1$ requires a compensating change in $m_\textrm{A}$ or $m_\textrm{B}$.
For the chosen incoming channel, resonances first-order in $\hat{H}_\textrm{Fc}$ are expected from two closed channels:
$(m^\textrm{res}_{f,\textrm{C}},m^\textrm{res}_\textrm{A},m^\textrm{res}_\textrm{B}) = (0,-3,1/2)$ and $(0,-4,3/2)$. Resonances from both these channels occur in the same region of 0 to 200 G, but there are spin-dependent shifts due to the diagonal portion of $\hat{H}_\textrm{Fc}$; depending on the magnitude of these, the states may be either close together or shifted far from each other within this region.

Additional resonances may occur due to closed channels with $\tilde{f}^\textrm{res}_\textrm{C}=2$. This state is excited by 6.83 GHz at zero field; to cause resonances accessible in magnetic fields below 600 G,
the states for $(\tilde{f}^\textrm{res}_\textrm{C},m^\textrm{res}_{f,\textrm{C}})=(2,1)$ would need to be bound by between 6.83 and 7.81 GHz.
The relative magnetic moment is larger in this case, but the states are much sparser in energy this far below threshold. It is therefore relatively unlikely (probability $\sim 19\%$) that resonances due to states with $\tilde{f}^\textrm{res}_\textrm{C}=2$ will exist at such fields.
This likelihood is considerably larger when the free atom has a substantially smaller hyperfine splitting, as for isotopes of Li, Na or K.
An example of such crossings is shown in Fig.\ \ref{fig:crossing}(b), showing that they will occur at low fields only if there is a state near the boundary between the fourth and fifth bins.
Resonances may also exist due to states in channels with $(\tilde{f}^\textrm{res}_\textrm{C},m^\textrm{res}_{f,\textrm{C}})=(2,0)$ or (2,2).
These resonances are analogous to Type I and II resonances that can occur in systems such as RbSr and CsYb \cite{Zuchowski:RbSr:2010, Brue:AlkYb:2013, Barbe:RbSr:2018, Yang:CsYb:2019}.

The different values of $\Delta m_{f,\textrm{C}}=0,\pm1$ require different combinations $(m^\textrm{res}_\textrm{A},m^\textrm{res}_\textrm{B})$ to produce resonances at each of the three crossings. For our chosen incoming state, for $\Delta m_{f,\textrm{C}}=-1$ we once again expect first-order resonances due to states with $(-3,1/2)$ and $(-4,3/2)$; for $\Delta m_{f,\textrm{C}}=0$ we expect $(-4,1/2)$; and for $\Delta m_{f,\textrm{C}}=+1$ we expect $(-4,-1/2)$.
The states that cause these resonances are more deeply bound than those shown in Fig.\ \ref{fig:crossing}(a), so their wavefunctions will have greater density at short range, sample more of the regions where the Fermi contact interaction is strong, and thus be spread out over a larger range of energy and field by spin-dependent shifts.

Further classes of resonances may exist in molecular collisions, due to closed channels with $n^\textrm{res}>0$ but $N^\textrm{res}=0$.
For KRb the $n=1$ state is located 2.23 GHz above $n=0$ \cite{Ospelkaus:hyperfine-control:2010}, and the $n=2$ state is located 6.68 GHz above. The molecule+atom threshold for $(n,f_\textrm{C})=(2,1)$ is thus below that for $(0,2)$ by just 150 MHz at zero field.
The states that can cause resonances at low fields must be bound by slightly more than the rotational splittings. This is similar to the case of $(0,2)$, but with smaller relative magnetic moments. This means that the energy windows needed to produce resonances below 600~G are narrower: for $n^\textrm{res}=1$, the states must be bound between 2.23 and 2.62 GHz, with a probability of 15\%; for $n^\textrm{res}=2$, the states must be bound between 6.68 and 7.07 GHz, with a probability of 8\%.
These states are coupled to the incoming channel by the part of the Fermi contact interaction that is anisotropic in the sense that it depends on the Jacobi coordinate $\theta$.
In Fig.\ \ref{fig:crossing}(c) we show examples of these states with (1,1) and (2,1), and $N^\textrm{res}=0$ such that $L^\textrm{res}=n^\textrm{res}$.
This shows that the states for $(2,1)$ will cause resonances only if they lie near the boundary between the fourth and fifth bins, as for $(0,2)$, and those for $(1,1)$ if they lie in the middle of the third bin.
The components $(m^\textrm{res}_\textrm{A},m^\textrm{res}_\textrm{B})$ that can cause resonances are the same as for $n^\textrm{res}=0$.

There can also be resonances from states supported by higher thresholds.
These include thresholds with $n^\textrm{res}>2$, or with $n^\textrm{res}>0$ and $\tilde{f}^\textrm{res}_\textrm{C}=2$. However these states must be even more deeply bound, so are more sparse. At sufficient depth, such states become increasingly strongly coupled to short-range states and are expected to become part of the chaotic manifold(s). No current theories exist to describe how and for what states this happens.

One special feature of this system occurs due to the near-degeneracy, mentioned above, between the rotational excitation to $n=2$ and the hyperfine splitting.
As seen in Fig.\ \ref{fig:crossing}(c), a bound state in the top bin for a channel with $(n,\tilde{f}_\textrm{C},m_{f,\textrm{C}})=(2,1,-1)$ crosses the threshold with $(0,2,-2)$ between 70 and 140 G, and the corresponding states in the second bin crosses between 140 and 580 G.
This will produce an additional type of resonance at thresholds with $\tilde{f}^\textrm{in}_\textrm{C}=2.$

Finally, there can be resonances due to states with $N^\textrm{res}>0$. These are coupled to the incoming state by only the couplings described in Sec.\ \ref{sec:off_diag}. We focus on states with
$(n^\textrm{res},\tilde{f}^\textrm{res}_\textrm{C},L^\textrm{res})=(0,1,1)$ and (0,1,2) (with $N^\textrm{res}=1$ and 2, respectively), which are those most likely to cause low-field resonances.
Examples of these states are shown in Fig.\ \ref{fig:crossing}(d). These are again shown for $a=\infty$, but for $L^\textrm{res}=1$ and 2 these states are not at the bin boundary. We therefore also show the bin boundaries as discussed in Sec.\ \ref{sec:bins}. This shows that such resonances are expected below 360 and 560 G for $L^\textrm{res}=1$ and 2, respectively.
There are many more components $(m_\textrm{A}^\textrm{res},m_\textrm{B}^\textrm{res})$ that can cause resonances in this case, because the sum $m_{f,\textrm{tot}}=m_\textrm{A}+m_\textrm{B}+m_{f,\textrm{C}}$ can change, compensated by $M_N$ to conserve $M_F$. The components that cause the strongest resonances will depend on which of the couplings discussed in Sec.\ \ref{sec:off_diag} are most important.
We have previously attributed the magnetic Feshbach resonances observed for $^{40}$K+NaK to states of this type \cite{Wang:K_NaK:2021}.

\section{Conclusions}
\label{sec:conc}

We have provided a framework for understanding long-range states of alkali-metal atom-diatom complexes, and the types of magnetic Feshbach resonances they can produce.
We first explored the terms that exist in the Hamiltonian for such triatomic systems and the ways in which they couple the six sources of angular momentum present. We separated the couplings into those that conserve the total spin-free angular momentum $N$ and those that can change it. The former category includes the electronic interaction potential and the Fermi-contact hyperfine interaction. The couplings in the latter category include spin-rotation, tensor hyperfine, and anisotropic Zeeman interactions. These terms are all expected to be significantly weaker than the Fermi-contact interaction, but they allow coupling to a wider variety of channels and thus may still be crucial to the dynamics.

We then considered the nature of the long-range states themselves. In contrast to the short-range states of the complex considered previously \cite{Mayle:2012, Frye:chaos:2016}, these are expected to retain much of the character of the separated atom and diatomic molecule. They are likely to have structured (non-chaotic) patterns of energy levels; the levels exist in `bins' of energy below each threshold that are determined by the reduced mass and the long-range potential.

Finally, through a case study of KRb+Rb, we considered the types of magnetic Feshbach resonances that can occur. These split into a number of broad categories: those which involve only Zeeman splittings, with no hyperfine or rotational excitation; those which involve hyperfine excitation but no rotation; those which involve rotational excitation (with or without hyperfine excitation) but conserve the total spin-free angular momentum; and those that change the spin-free angular momentum. For each type we considered the ranges of field where they occur and the conditions that need to be met for the resonances to exist at experimentally accessible fields.

The work described here paves the way for future studies of Feshbach resonances in alkali-metal atom--diatom systems and for their use to control collisions and form ultracold triatomic molecules.

\begin{acknowledgments}
This work was supported by the U.K. Engineering and Physical Sciences Research Council (EPSRC)
Grants No.\ EP/P01058X/1 and EP/W00299X/1.
\end{acknowledgments}

\appendix
\section{Channel functions}

The channel functions used in the coupled-channel expansion (\ref{eq:exp}) are based on, but not equivalent to, the eigenfunctions of the separated atom and molecule. They are
\begin{equation}
\Phi_j(\xi) = \psi^\textrm{C}_{\tilde{f},m_{f,\textrm{C}}} \psi^\textrm{AB}_{m_\textrm{A},m_\textrm{B}} \psi^\textrm{ABC}_{vnLNM_N}(\vec{r},\hat{R}).
\label{eq:basis-N}
\end{equation}
Here $\psi^\textrm{C}_{\tilde{f},m_{f,\textrm{C}}}$ is an eigenstate of the free atom in a magnetic field, $\psi^\textrm{AB}_{m_\textrm{A},m_\textrm{B}}$ is a spin state of the free diatomic molecule, and the coupled spatial function is
\begin{align}
\psi^\textrm{ABC}_{vnLNM_N}(\vec{r},\hat{R}) &=  \sum_{m_n,M_L} \langle n m_n L M_L | N M_N \rangle
\nonumber\\ &\times \psi_{vn}(r) Y_{nm_n}(\beta_r,\alpha_r) Y_{LM_L}(\beta_R,\alpha_R),
\label{eq:nLN}
\end{align}
where $\psi_{vn}(r)$ is a vibrational function for the diatomic molecule, the functions $Y$ are spherical harmonics whose arguments are the polar coordinates of unit vectors $\hat{r}$ and $\hat{R}$ along $\vec{r}$ and $\vec{R}$, and $\langle n m_n L M_L | N M_N \rangle$ is a Clebsch-Gordan coefficient. These channel functions are eigenfunctions of the Hamiltonian of the free atom and the vibration-rotation and nuclear Zeeman terms in the Hamiltonian of the free diatomic molecule. However, there are small off-diagonal terms arising from the hyperfine, rotational Zeeman and ac Stark terms for the free molecule, and from the interaction operator $\hat{V}(R,\xi)$.

\bibliographystyle{long_bib}
\bibliography{../all,atom-molecule-long-range}

\end{document}